\documentclass[a4paper,preprint,prd,showpacs,preprintnumbers,amsmath,amssymb,floatfix]{revtex4}
\usepackage{amsmath}
\usepackage{graphicx} 
\usepackage{dcolumn} 
\usepackage{bm} 
\usepackage{ulem} 
\usepackage[usenames]{color}
\usepackage{epstopdf}
\usepackage{float}
\usepackage{subcaption} 
\usepackage{multirow}
\usepackage{array}
\usepackage{setspace}
\usepackage{caption}
\usepackage{hyperref}
\usepackage{rotating}
\usepackage{xcolor} 
\usepackage{ulem}

\graphicspath{{fig/}} 

\newcommand{\nc}{\newcommand} 
\nc{\vc}[1]{\mbox{\boldmath $#1$}} 
\nc{\del}{\partial} 
\nc{\bra}{\langle} 
\nc{\ket}{\rangle} 
\nc{\bras}[1]{\langle #1|} 
\nc{\kets}[1]{|#1\rangle} 
\nc{\mapleft}[1]{\smash{\mathop{\,\hbox to 1.5cm{\rightarrowfill}\,}\limits_{#1}}}
\nc{\nn}{\\\nonumber}
\nc{\vs}{\vspace{-0.275cm}}
\nc{\fra}{\frac{1}{2}}
\nc{\mb}{\mathbf}

\begin{document}

	\preprint{}
\title{
	Radial oscillations of neutron stars within density-dependent relativistic-mean field model} 

\author{ Xuesong Geng}
\affiliation{School of Physics, Nankai University, Tianjin 300071,  China}
\author{ Kaixuan Huang}
\affiliation{School of Physics, Nankai University, Tianjin 300071,  China}
\author{Hong Shen}
\email{songtc@nankai.edu.cn}
\affiliation{School of Physics, Nankai University, Tianjin 300071,  China}
\author{Lei Li}
\email{lilei@nankai.edu.cn}
\affiliation{School of Physics, Nankai University, Tianjin 300071,  China}
\author{Jinniu Hu}
\email{hujinniu@nankai.edu.cn}
\affiliation{School of Physics, Nankai University, Tianjin 300071,  China}
\affiliation{Shenzhen Research Institute of Nankai University, Shenzhen 518083, China}
\date{\today} 	
	
\title{Radial oscillations of neutron stars within density-dependent relativistic-mean field model\thanks{Supported by in part by the National Natural Science Foundation of China (Grants  Nos. 12175109, 12475149), the Natural Science Foundation of Guangdong Province (Grant  No: 2024A1515010911)}}

\begin{abstract}
The radial oscillations of neutron stars are studied using equations of state derived from density-dependent relativistic mean-field (DDRMF) models, which effectively describe the ground-state properties of finite nuclei. A novel numerical approach, the finite volume method (FVM), is employed to solve the eigenvalue problem associated with oscillation frequencies. Compared to conventional methods such as the finite difference method and shooting method, the FVM avoids the numerical instability encountered at high frequencies with an equation of state that includes a discontinuous adiabatic index and offers greater computational efficiency. The oscillation frequencies of high-order modes exhibit a similar trend of change. The radial displacements and pressure perturbations are largely influenced by the EOSs of crust region. {The frequency of the first excited state shows a strong linear relationship with both the slope and skewness parameters of the symmetry energy.} These findings suggest that the density dependence of the symmetry energy can be constrained through observations of neutron star radial oscillation frequencies.
\end{abstract}


\maketitle

\section{Introduction}
Neutron stars (NSs) are stellar remnants formed from the supernova explosions of massive stars, with central densities several times greater than the nuclear saturation density. They serve as natural laboratories for investigating the equation of state (EOS) of dense matter under extreme conditions, providing insights into quantum chromodynamics (QCD) and testing the validity of modern nuclear many-body frameworks \citep{Glendenning1997}.
	
Astronomical observation technologies for NSs have undergone significant improvements over the past decades. The successful operation of gravitational wave (GW) detectors by the LIGO Scientific Collaboration and Virgo Collaboration (LVC) in 2015 marked the advent of multi-messenger astronomy with GWs. Subsequently, the detection of the GW170817 event—the first binary neutron star (BNS) merger—by LVC provided an estimate of the tidal deformability for NSs around $1.4M_\odot$, introducing a new constraint on the EOS of dense matter.
	
Most observed NS masses are approximately $1.4M_\odot$, although some exceed $2M_\odot$. Precise mass measurements of massive pulsars, such as PSR J1614-2230 \citep{Arzoumanian2018}, PSR J0348+0432 \citep{Antoniadis2013}, and PSR J0740+6620 \citep{Fonseca2021}, using the Shapiro effect, require the predicted maximum NS mass to surpass $2M_\odot$. Recent simultaneous measurements of the masses and radii of PSR J0030+0451 \citep{Riley2019,Miller2019}, PSR J0740+6620 \citep{Riley2021,Miller2021}, PSR J0437-4715 \citep{Reardon2024,Choudhury2024}, and PSR J1231-1411 \citep{Salmi2024} by the Neutron Star Interior Composition Explorer (NICER) have provided further constraints on EOS behavior.
	
Understanding the internal composition of NSs is crucial for constructing their EOS. Oscillations of NSs offer a valuable probe for exploring their internal structures, akin to the role of asteroseismology in studying ordinary stars. These oscillations are categorized into radial and non-radial types, with the latter further divided into modes based on restoring forces \citep{Kokkotas1999}. Radial oscillations, as the simplest type, involve expansions and contractions around the star's equilibrium structure while preserving its spherical symmetry \citep{Vaeth1992,Kokkotas2001,Passamonti2005,Passamonti2006,Brillante2014,Panotopoulos2017,Flores2017,Lugones2010}.
	
Although radial oscillations do not directly emit gravitational waves, they can interact with non-radial oscillations, amplifying gravitational wave signals \citep{Passamonti2006}. These interactions may be detectable with next-generation detectors like the Cosmic Explorer \citep{Evans2021} and the Einstein Telescope \citep{Punturo2010}. Furthermore, short gamma-ray bursts (SGRBs) generated by binary NS mergers are influenced by radial oscillations \citep{Chirenti2019}. Consequently, recent studies have examined radial oscillations in NSs considering various factors such as strangeness, quark-hadron phase transitions, dark matter, and proto-neutron stars \citep{Glass1983,Panotopoulos2017,Flores2017,Panotopoulos2018,Sagun2020,Jim2021,Li2022,hong2022,Sen2023,Routaray2023,Rather2023,sun2024}.

The EOS of NSs can be derived from nuclear density functional theories, where nucleon–nucleon (NN) interactions are determined by fitting ground-state properties of finite nuclei or empirical saturation properties of infinite nuclear matter \citep{Vautherin1972,Shen1998,Shen2002,Douchin2001,Bao2014,Bao2014a}. The relativistic mean-field (RMF) model, introduced by Walecka using the Hartree approximation \citep{Walecka1974}, has achieved great success in nuclear physics and astrophysical studies. Enhancements through nonlinear and coupled terms enable the RMF model to accurately describe the ground-state properties of most nuclei \citep{Boguta1977,Serot1979,Sugahara1994,Horowitz2001}. The density-dependent RMF (DDRMF) and relativistic Hartree-Fock (DDRHF) models incorporate density-dependent meson-nucleon couplings to account for medium effects \citep{Brockmann1992}, making DDRMF parameterizations widely applicable for NS studies. Certain parameter sets accurately describe both finite nuclei and extreme NS configurations \citep{Huang2020,huang2024}.
	
In this work, we systematically study the radial oscillations of NSs using EOSs derived from DDRMF parameterizations, which provide accurate ground-state properties for finite nuclei. Most existing numerical methods for solving radial oscillation equations rely on the shooting method, with limited applications of the finite difference method (FDM) \citep{Vaeth1992,Kokkotas2001,Barta2021,Glass1983}. However, if the conservation form of the differential equation is not taken into account, the FDM fails to handle discontinuities in the adiabatic index of the EOS. Therefore, we discuss this issue and introduce the finite volume method (FVM), which satisfies the conservation property. The radial oscillation equations, FVM, and DDRMF models are introduced in Section ~II. The results and discussions are presented in Section ~III. The conclusions are summarized in Section ~IV.

\section{THEORETICAL FRAMEWORK}
	\label{II}
	\subsection{The radial oscillations of neutron stars}
	A static NS can be regarded as a sphere because of the massive gravitational field, described by the Schwarzschild metric \citep{Chanmugam1977}
	\begin{equation}
		\mathrm{d}s^{2}=e^{\nu}\left(c\mathrm{d}t\right)^{2}-e^{\lambda 	}\left(\mathrm{d}r\right)^{2}-r^{2}\left(\mathrm{d}\theta ^{2}+\sin^{2}\theta \mathrm{d}\phi ^{2}\right),
		\label{metric}
	\end{equation}
	where $e^{\lambda}$ and $e^{\nu}$ are the metric functions.
	
	The Tolman-Oppenheimer-Volkoff(TOV) equations describing the mass and radius of NS \citep{Tolman1939,Oppenheimer1939} can be obtained by solving the Einstein field equation with the above-defined metric as 
	
		\begin{align}
			\frac{\mathrm{d}P}{\mathrm{d} r}&=-\frac{Gm}{c^{2}r^{2}}\frac{\left(P+\varepsilon \right)\left(1+\frac{4\pi r^{3}P}{mc^{2}} \right)}{\left(1-\frac{2Gm}{c^{2}r} \right)},  \nonumber\\
			\frac{\mathrm{d} m}{\mathrm{d} r}&=\frac{4\pi r^{2}\varepsilon }{c^{2}},  
			\label{equ-TOV} 
		\end{align}
	
	where the corresponding metric functions, $\lambda(r)$ and $\nu(r)$ express as
	\begin{equation}
		e^{\lambda(r)}=\left(1- \frac{2Gm}{rc^{2}} \right)^{-1},~~~\frac{\mathrm{d} \nu}{\mathrm{d} r}=- \frac{2}{P+\varepsilon } \frac{\mathrm{d} P}{\mathrm{d} r}, 
	\end{equation}
	and they are in the surface of NS related with the total mass, $M$ and radius $R$
	\begin{equation}
		e^{\nu(R)}=e^{-\lambda(R)}=\left(1-\frac{2Gm}{Rc^{2}} \right).
	\end{equation}
	
	After giving a EOS of NS matter, i.e., $P-\varepsilon$ relation, the TOV equations as first-order ordinary differential equations can be solved with initial conditions $m(r=0)=0$, $P(r=0)=P_{c}$, and the boundary condition $P(r=R)=0$ at surface, $R$, where $P_{c}$ is the central pressure of NS. 
	
	Considering a spherically symmetric system only with a small perturbative radial movement,
	\begin{equation}
		\delta r(r,t)=\Delta r(r)e^{i\omega t},
		\label{Delta_r}
	\end{equation}
	the metric Eq. \eqref{metric} is no longer time independent. The radial oscillation properties can be obtained from the Einstein field equation based on the static equilibrium structure \citep{Bardeen1966}. One can define the small perturbations of the dimensionless quantities $\xi=\Delta r/r$ and $\eta=\Delta P/P$, where $\Delta r$ and $\Delta P$ represent the radial displacement and the pressure perturbation. They are governed by \citep{Chandrasekhar1964,Chandrasekhar1964apj,Chanmugam1977}
    \begin{align}
			\frac{\mathrm{d} \xi}{\mathrm{d} r}=&-\left[\frac{3}{r}+\frac{\mathrm{d} p}{\mathrm{d} r}\frac{1}{(P+\varepsilon)}\right]\xi-\frac{\eta}{r\Gamma}, \nonumber\\ 
			\frac{\mathrm{d} \eta}{\mathrm{d} r} =& \biggl[	\frac{\omega^{2}}{c^{2}}e^{\lambda-\nu}\frac{\left(P+\varepsilon\right)}{P}r-\frac{4}{P}\frac{\mathrm{d} P}{\mathrm{d} r}-\frac{8\pi G}{c^{4}}e^{\lambda}\left(P+\varepsilon\right)r \nonumber\\
			&+\left(\frac{\mathrm{d} P}{\mathrm{d} r} \right)^{2}\frac{r}{P\left(P+\varepsilon\right)}\biggr]\xi  \nonumber\\
            &-\left[\frac{\mathrm{d} P}{\mathrm{d} r}\frac{\varepsilon}{P\left(P+\varepsilon\right)}+\frac{4\pi G}{c^{4}}\left(P+\varepsilon\right)e^{\lambda}r \right]\eta ,
			\label{roeq}
    \end{align}
	where $\omega$ is the eigenfrequency of radial oscillation and $\Gamma$ is the adiabatic index of NS matter
	\begin{equation}
		\Gamma =\left(1+\frac{\varepsilon}{P} \right)c_{s}^{2},
	\end{equation} 
	where $c_{s}$ is the speed of sound.
	
	In addition, $\xi$ and $\eta$ in Eq. \eqref{roeq} satisfy two boundary conditions. At the center region, $r=0$
	\begin{equation}
		3\Gamma \xi_{r=0}+\eta_{r=0}=0,
		\label{equ-boundary1}
	\end{equation}
	and at the surface of NS, $r=R$
	\begin{equation}
		\eta_{R}=-\left[4+\left(1-2\frac{GM}{Rc^{2}}\right)^{-1}\left(\frac{GM}{Rc^{2}}+\frac{\omega ^{2}R^{3}}{GM}\right)\right]\xi_{R}.
		\label{equ-boundary2}
	\end{equation}
	
	A new quantity can be defined to present the perturbative movement as, $\zeta=r^{2}e^{-\nu}\Delta r$. The Eq. \eqref{roeq} can be transferred to  a linearized radial perturbation equations \citep{Kokkotas2001} as
	\begin{equation}
		\frac{\mathrm{d}}{\mathrm{d} r}\left(H \frac{\mathrm{d} \zeta}{\mathrm{d} r}\right)+\left(\omega^{2} W+Q\right) \zeta  = 0
		\label{ode}
	\end{equation}
	with 
    \begin{align}
			H & = r^{-2}\left(P+\varepsilon \right) e^{\frac{1}{2}\lambda+\frac{3}{2} \nu} 	c_{s}^{2},\nonumber\\
			W & = r^{-2}\left(P+\varepsilon \right) e^{\frac{3}{2} \lambda+\frac{1}{2}\nu}, 	\nonumber\\
			Q & = r^{-2}\left(P+\varepsilon \right) e^{\frac{1}{2}\lambda+\frac{3}{2} \nu}\left(\frac{1}{4}\nu^{\prime 2}+\frac{2}{r} \nu^{\prime}-\frac{8 \pi G}{c^{4}} e^{ \lambda} P\right).
    \end{align}
	The boundary conditions are
	\begin{equation}
		\begin{aligned}
			\Delta r\left(r=0\right)=0,\ \ \ \Delta P\left(r=R\right)=0,
		\end{aligned}
	\end{equation}
	where $\Delta P$ is Lagrangian variation of the pressure defined as
	\begin{equation}
		\Delta P=-r^{-2} e^{\frac{1}{2}\nu}\left(P+\varepsilon \right)c^{2}_{s}\zeta'.
	\end{equation}
	
	The eigenvalues $\omega^{2}_{n}$ in Eq. \eqref{ode} are discrete real numbers, which follow
	\begin{equation}
		\omega ^{2}_{0}<\omega ^{2}_{1}<\cdots <\omega ^{2}_{n}.
	\end{equation}
	For the imaginary solution, according to Eq. \eqref{Delta_r}, the star will become unstable \citep{Harrison1965}, while oscillations will be harmonic and stable for real $\omega$. When the central density exceeds the critical density corresponding to the maximum mass, $\omega_{0}$ will become imaginary. The radial oscillation frequency is defined as 
	\begin{equation}
		\nu_{n}=\frac{\omega_n}{2\pi}.
	\end{equation}

	\subsection{The finite volume method}
	{The FDM} is a discretized scheme to treat the Sturm-Liouville eigenvalue equation such as Eq. \eqref{ode}, which requires the eigenfunction changes smoothly. However, due to crust-core phase transition, the adiabatic index, $\Gamma$ has a {discontinuity} at crust region. Once the crust is considered, the eigenvalues obtained by FDM have the obvious differences with those from shooting method. It was found that the finite volume method (FVM), which is applied to solve the heat transfer equation \citep{Cebula2014} is a good tool to deal with such problem. We found that in the presence of a discontinuous adiabatic index, it is essential to carefully consider the conservative form of the differential equation—an issue that was not addressed in previous studies \citep{Vaeth1992,Kokkotas2001,Barta2021,Glass1983}. Similar to the FDM, a mesh is constructed in FVM. The differential equations form an equilibrium equation at each element of the grid.  Using the integral form of the equilibrium equation and the midpoint theorem, a discrete equation in the coordinate space can be obtained.
	
	After solving the TOV equation, the radius in the eigenvalue equation,  Eq. \eqref{ode}, is divided into equal lattices $N$ to get the step, $h$
	\begin{equation}
		r_{i}=i\frac{R}{N}=ih.
	\end{equation}
	Other quantities with the lower corner label $r_{i}$ represent their corresponding values at $r_{i}$. One can integrate Eq. \eqref{ode} over the discrete cells between $[r_{i-1/2},~r_{i+1/2}]$,
	\begin{equation}
		\int_{r_{i-1/2}}^{r_{i+1/2}} \frac{\mathrm{d}}{\mathrm{d} r}\left(H \frac{\mathrm{d} \zeta}{\mathrm{d} r}\right)\mathrm{d}r
		+\int_{r_{i-1/2}}^{r_{i+1/2}} \left [   \left(\omega^{2} W+Q\right) \zeta \right ]\mathrm{d}r = 0,
	\end{equation}
	where the first part can be integrated analytically,
	\begin{equation}
		\int_{r_{i-1/2}}^{r_{i+1/2}} \frac{\mathrm{d}}{\mathrm{d} r}\left(H \frac{\mathrm{d} \zeta}{\mathrm{d} r}\right)\mathrm{d}r
		=\left(H \frac{\mathrm{d} \zeta}{\mathrm{d} r}\right)_{r_{i+1/2}}-\left(H \frac{\mathrm{d} \zeta}{\mathrm{d} r}\right)_{r_{i-1/2}}.
	\end{equation}
	$H \frac{\mathrm{d} \zeta}{\mathrm{d} r}$ is rewritten as,
	\begin{equation}
		\frac{\mathrm{d} \zeta}{\mathrm{d} r}=\frac{H \frac{\mathrm{d} \zeta}{\mathrm{d} r}}{H} ,
	\end{equation}
	We integrate it in the interval $\left [r_{i-1}, r_{i}  \right ] $, and take an approximation,
	\begin{equation}
		\zeta_{r_{i}}-\zeta_{r_{i-1}}=\int_{r_{i-1}}^{r_{i}} \frac{H \frac{\mathrm{d} \zeta}{\mathrm{d} r}}{H} \mathrm{d}r\approx  
		\left( H \frac{\mathrm{d} \zeta}{\mathrm{d} r}\right)_{r_{i-1/2}} \int_{r_{i-1}}^{r_{i}} \frac{1}{H} \mathrm{d}r,
	\end{equation}
	so 
	\begin{equation}
		\left( H \frac{\mathrm{d} \zeta}{\mathrm{d} r}\right)_{r_{i-1/2}}\approx  \frac{\zeta_{r_{i}}-\zeta_{r_{i-1}}}{\int_{r_{i-1}}^{r_{i}} \frac{1}{H} \mathrm{d}r}   ,
	\end{equation}
	where $\int_{r_{i-1}}^{r_{i}} \frac{1}{H} \mathrm{d}r$ can be written approximately to $\frac{r_{i}-r_{i-1}}{H_{r_{i-1/2}}}$. In the same way, we can obtain $\left( H \frac{\mathrm{d} \zeta}{\mathrm{d} r}\right)_{r_{i+1/2}}$. Meanwhile the other parts can be expressed as
	\begin{align}
			\int_{r_{i-1/2}}^{r_{i+1/2}} \left [   \left(\omega^{2} W+Q\right) \zeta \right ]\mathrm{d}r &\approx \zeta_{r_{i}} \int_{r_{i-1/2}}^{r_{i+1/2}} \left(\omega^{2} W+Q\right) \mathrm{d}r \nonumber\\
			&\approx \zeta_{r_{i}}\left(r_{i+1/2}-r_{i-1/2})(\omega^{2}W_{r_{i}}+Q_{r_{i}}\right).
			\label{approx1}
	\end{align}
	
	For a uniform grid, $r_{i}-r_{i-1}=h$, by combining above equations, one can get a discretized equation in lattice space, 
	\begin{align}\label{odemat}
	H_{r_{i+1/2}}\frac{\left(\zeta_{r_{i+1}}-\zeta_{r_{i}}\right)}{h} &-H_{r_{i-1/2}}\frac{\left(\zeta_{r_{i}}-\zeta_{r_{i-1}}\right)}{h} \nonumber\\
            &+h\left(\omega^{2}W_{r_{i}}+Q_{r_{i}}\right)\zeta_{r_{i}}=0.
	\end{align}
   which can also be obtained by calculating the derivative of the  term $H\zeta'$ in a  staggered grid  in FDM.
    
	There are two boundary conditions at $r=0$ and $r=R$,
	\begin{equation}
		\zeta_{r_{0}}=0,\ \ \  \zeta_{r_{N+1}}=\zeta_{r_{N-1}}.
	\end{equation}
	A matrix form can be generated with Eq. \eqref{odemat},
	\begin{equation}
		\left(A-\omega^{2}_{n}I \right)\zeta=0,
	\end{equation}
	where $A$ is the tridiagonal matrix with the coefficients,
	
\small{
    \begin{align}
        \setlength{\arraycolsep}{0.5pt}
		A = \begin{pmatrix}
			a_{1,1} & a_{1,2} & 0 & \cdots & 0 & 0 & 0 \\
			a_{2,1} & a_{2,2} & a_{2,3} & \cdots & 0 & 0 & 0 \\
			0 & a_{3,2} & a_{3,3} & \ddots & \vdots & 0 & 0 \\
			\vdots & \vdots & \ddots & \ddots & a_{N-2,N-3} & 0 & 0 \\
			0 & 0 & \cdots & a_{N-2,N-3} & a_{N-2,N-2} & a_{N-2,N-1} & 0 \\
			0 & 0 & \cdots & 0 & a_{N-1,N-2} & a_{N-1,N-1} & a_{N-1,N} \\
			0 & 0 & \cdots & 0 & 0 & a_{N,N-1} & a_{N,N}
	{multicols}	\end{pmatrix}
	\end{align}
	where
	\begin{align*}
		a_{i,i} &= -\frac{Q_{r_i}}{W_{r_i}} + \frac{H_{r_{i+1/2}} + H_{r_{i-1/2}}}{h^2 W_{r_i}}, \quad i = 1, 2, \ldots, N-1, \\ 
		a_{i,i+1} &= -\frac{H_{r_{i+1/2}}}{h^2 W_{r_i}}, \quad i = 1, 2, \ldots, N-1, \\
		a_{i,i-1} &= -\frac{H_{r_{i-1/2}}}{h^2 W_{r_{i}}}, \quad i =  2,3, \ldots, N-1, \\ 
		a_{N,N-1} &= -\frac{H_{r_{N-1/2}} + H_{r_N}}{h^2 W_{r_N}}, 
		a_{N,N}= -\frac{Q_{r_N}}{W_{r_N}} + \frac{H_{r_{N-1/2}} + H_{r_N}}{h^2 W_{r_N}},
	\end{align*}  
    }
	and $H_{r_{N+1/2}}$ is approximated to $H_{r_{N}}$.

   The FDM obtains the discretized equation by approximating the first and second derivatives of $\zeta$ using central differences. One advantage of Eq. \eqref{odemat} is that it preserves the conservation form of the differential equation while allowing discontinuities in the coefficients. Additionally, a non-uniform grid can be adopted, and when integrating the coefficients, one may employ the rectangle rule or Simpson’s rule. These strategies enhance the capability of coarse grids to handle oscillatory equations.
	
	\subsection{Density-dependent relativistic mean-field model}
	In our work, the EOSs to solve the radial oscillation equation of NS are generated by the DDRMF model. It is an effective field theory of interacting mesons and nucleons, which takes into account the nucleon-nucleon interaction in the dense matter affected by the nuclear medium.  We adopted $14$ groups of parameters of the DDRMF model,  DDLZ1\citep{wei20}, DDMEX\citep{taninah20}, DDMEX1\citep{taninah23}, DDMEX2\citep{taninah23}, DDMEXY\citep{taninah23}, DDME2\citep{lalazissis05}, DDME1\citep{niksic02}, DD2\citep{typel10}, PKDD\citep{long04}, TW99\citep{typel99}, DDV\citep{typel20}, DDVT\citep{typel20}, DDVTD\citep{typel20}, DDT\citep{typel2024}, which can precisely describe the ground-state properties of finite nuclei. The maximum NS mass and corresponding radius from most of these parameters have been confirmed to meet the astronomical observation data \citep{Huang2020}. The theoretical framework of DDRMF model was given in detail at Ref. \citep{huang2024arxiv}. The Lagrangian of the DDRMF model can be expressed as,
        \begin{align}
            \mathcal{L}_{\rm DD}
            =&\sum_{N=n,p}\overline{\psi}_ N\left[\gamma^{\mu}\left(i\partial_{\mu}-\Gamma_{\omega N}(\rho_ N)\omega_{\mu}-\frac{\Gamma_{\rho N}(\rho_N)}{2}\vec{\rho}_{\mu}\vec{\tau}\right) \right. \notag\\ 
            &\phantom{\bigg[}-\left(M_ N-\Gamma_{\sigma N}(\rho_N)\sigma-\Gamma_{\delta N}(\rho_{N})\vec{\delta}\vec{\tau}  \right)\bigg]\psi_N \nonumber\\
            &+\frac{1}{2}\left(\partial^{\mu}\sigma\partial_{\mu}\sigma-m_{\sigma}^2\sigma^2\right)
            +\frac{1}{2}\left(\partial^{\mu}\vec{\delta}\partial_{\mu}\vec{\delta}-m_{\delta}^2\vec{\delta}^2\right)\nonumber\\
            &-\frac{1}{4}W^{\mu\nu}W_{\mu\nu}+\frac{1}{2}m_{\omega}^2\omega_{\mu}\omega^{\mu}-\frac{1}{4}\vec{R}^{\mu\nu}\vec{R}_{\mu\nu}+\frac{1}{2}m_{\rho}^2\vec{\rho}_{\mu}\vec{\rho}^{\mu},
        \end{align}
    
		where $W_{\mu\nu}=\partial_\mu\omega_\nu-\partial_\nu\omega_\mu$ and $\vec{R}_{\mu\nu}=\partial_\mu\vec\rho_\nu-\partial_\nu\vec\rho_\mu$ are the asymmetric tensor fields. With the mean-field  and no-sea approximations, the energy density, $\varepsilon$ , and pressure, $P$, of infinite nuclear matter in DDRMF model can be written as  
		\begin{align}
			\varepsilon=&\frac{1}{2}m_{\sigma}^2\sigma^2+\frac{1}{2}m_{\delta}^2\delta^2-\frac{1}{2}m_{\omega}^2\omega^2-\frac{1}{2}m_{\rho}^2\rho^2 +\Gamma_{\omega  N}(\rho_N)\omega\rho_N \notag\\ &+\frac{\Gamma_{\rho N}(\rho_N)}{2}\rho\rho_3+\mathcal{E}_{\rm kin}^p+\mathcal{E}_{\rm kin}^n,\notag\\
			P=&\rho_N\Sigma_{R}(\rho_N)-\frac{1}{2}m_{\sigma}^2\sigma^2-\frac{1}{2}m_{\delta}^2\delta^2+\frac{1}{2}m_{\omega}^2\omega^2
            \notag\\&+\frac{1}{2}m_{\rho}^2\rho^2+P_{\rm kin}^p+P_{\rm kin}^n.
		\end{align} 
		Here, the contributions from the kinetic energy are 
		\begin{equation}
			\begin{aligned}
				\mathcal{E}_{\rm kin}^i&=\frac{\gamma}{2\pi^2}\int_{0}^{k_{Fi}}k^2\sqrt{k^2+{M_i^{*}}^{2}}dk
                \\&=\frac{\gamma}{16\pi^2}\left[k_{Fi}E_{Fi}^{*}\left(2k_{Fi}^2+{M_i^{*}}^2\right)+{M_i^{*}}^4{\rm ln}\frac{M_i^{*}}{k_{Fi}+E_{Fi}^{*}}\right], \\
				P_{\rm kin}^i&=\frac{\gamma}{6\pi^2}\int_{0}^{k_{Fi}}\frac{k^4 dk}{\sqrt{k^2+{M_i^{*}}^{2}}}
				\\&=\frac{\gamma}{48\pi^2}\left[k_{Fi}\left(2k_{Fi}^2-3{M_i^{*}}^2\right)E_{Fi}^{*}+3{M_i^{*}}^{4}{\rm ln}\frac{k_{Fi}+E_{Fi}^{*}}{M_i^{*}}\right].
			\end{aligned}
		\end{equation} 
		$\gamma=2$ is the spin degeneracy factor.  
		
		The binding energy per nucleon of the isospin asymmetric nuclear matter, $E/A=\varepsilon/\rho_N-M_N$, can be expanded in terms of the isospin asymmetry,  $\delta=\left(\rho_n-\rho_p\right) /\rho_N$, as
		\begin{gather}
			\frac{E}{A}(\rho_N, \delta)=\frac{E_{\rm SNM}}{A}(\rho_N)+E_{\mathrm{sym}}(\rho_N) \delta^2 +\mathcal{O}\left(\delta^4\right),
		\end{gather} 
		where $\rho_N=\rho_n+\rho_p$ is the baryon density with $\rho_n$ and $\rho_p$ denoting the neutron and proton densities, respectively. The first term $E_{\rm SNM}/A$ is the binding energy per nucleon of the symmetric nuclear matter. The symmetry energy $E_{\text {sym}}(\rho_N)$ can be expressed as
		\begin{equation} 
			E_{\rm sym}(\rho_N) =\left.\frac{1}{2!} \frac{\partial^2 E/A(\rho_N, \delta)}{\partial \delta^2}\right|_{\delta=0}.
		\end{equation}
		The odd-order terms in $\delta$ is attributed to the exchange symmetry between protons and neutrons in nuclear matter. The high-order terms are usually very small so we neglect them.  
		
		Around the normal nuclear matter saturation density $\rho_0$, $E_{\rm SNM}/A$ can be expanded, for example, up to fourth-order in terms of $\chi=\frac{\rho_N-\rho_0}{3\rho_0}$,
		\begin{equation}
			\frac{E_{\rm SNM}}{A}(\rho_N)=\frac{E}{A}+\frac{K}{2!} \chi^2+\frac{Q}{3!} \chi^3+\mathcal{O}\left(\chi^4\right).
		\end{equation} 
		The first term $E/A$ is the binding energy per nucleon at the saturation point. The curvature parameter $K$ is the imcompressibility and $Q$ is the skewness coefficients at $\rho_0$. They can be written as
		\begin{align}\label{eq.KQJe}
				K & =\left.9 \rho_0^2 \frac{\partial^2 E_{\rm SNM}/A}{\partial \rho_N^2}\right|_{\rho_N=\rho_0} =\left. 9\frac{\partial P}{\partial\rho_N}\right|_{\rho_N=\rho_0}, \notag\\
				Q & =\left.27 \rho_0^3 \frac{\partial^3 E_{\rm SNM}/A}{\partial \rho_N^3}\right|_{\rho_N=\rho_0}.
		\end{align}
		
		The NS is a extreme isospin asymmetry system, where the symmetry energy and its density-dependence play important roles.  Similarly, the symmetry energy $E_{\text {sym}}(\rho_N)$ can also be expanded around $\rho_0$ as   
		\begin{equation}
			\begin{aligned}
				E_{\text {sym}}(\rho_N)= & E_{\rm sym}+L \chi+\frac{K_{\text {sym }}}{2!} \chi^2 
				+\frac{Q_{\text {sym }}}{3!} \chi^3+\mathcal{O}\left(\chi^4\right),
			\end{aligned}
		\end{equation}
		where the $L,~K_{\text {sym }},~Q_{\text {sym }}$ are the slope parameter, curvature parameter, and skewness coefficients of the $E_{\rm sym}(\rho)$ at $\rho_0$, respectively, whose definitions are similar to Eqs. \ref{eq.KQJe},
		\begin{align} 
			L&= \left.3\rho_0\left(\frac{\partial E_{\rm sym}}{\partial\rho_B}\right)\right|_{\rho_N=\rho_0} ,\notag\\
			K_{\rm sym}&= \left.9\rho_0^2\frac{\partial^2E_{\rm sym}}{\partial\rho_B^2}\right|_{\rho_N=\rho_0},\notag\\
			Q_{\rm sym}&= \left.27\rho_0^3\frac{\partial^3E_{\rm sym}}{\partial\rho_B^3}\right|_{\rho_N=\rho_0}.
		\end{align}  
		
		The saturation properties of symmetric nuclear matter with different DDRMF effective interactions are calculated to investigate the relationship between saturation properties and radial oscillation frequencies, which are listed in Table~\ref{table.sat}, i.e., the saturation density, $\rho_0$, the binding energy per nucleon, $E/A$, incompressibility, $K$, skewness of the binding energy per nucleon, $Q$, the symmetry energy, $E_\text{sym}$, the slope of symmetry energy, $L$, the curvature and skewness of the symmetry energy, $K_{\rm sym},~Q_{\rm sym}$, and the effective neutron and proton masses, $M^*_n$ and $M^*_p$. 
		Their binding energy per nucleon, saturation density, and incompressibility are almost same. The symmetry energies and their slopes are consistent with each other except those from DDMEX2 and PKDD sets, whose $Q_{\rm sym}$ are very small. The effective nucleon masses from DDVT, DDVTD, and DDT are obviously larger than another ones. The tensor coupling between $\omega$ meson and nucleon are considered in these three sets.

\begin{table}[htb]
		\centering
            \scriptsize
            	\caption{Nuclear matter properties at saturation density generated by DDRMF parameterizations.}
            \label{table.sat}
                    \begin{tabular}{r|ccccccccccccccc}
				\hline\hline
				&$\rho_{0}[\rm fm^{-3}]$ ~&$E/A[\rm MeV]$ ~&$K[\rm MeV]$ ~&$Q[\rm MeV]$  ~&$E_{\rm sym}[\rm MeV]$  ~&$L[\rm MeV]$ ~&$K_{\rm sym}[\rm MeV]$ ~&$Q_{\rm sym}[\rm MeV]$ ~&$M_n^{*}/M_N$ ~&$M_p^{*}/M_N$\\
				\hline 
				DDLZ1 ~&0.1581 ~&-16.0584~&230.3336~&1331.7016 ~&32.1895 ~&42.1091 ~& -21.0259 ~& 919.4752 ~&0.5581 ~&0.5581  \\
				DDMEX ~&0.1516 ~&-16.1113~&267.1415~& 859.7477 ~&32.3639 ~&49.9555 ~& -73.6292 ~& 793.8295 ~&0.5554 ~&0.5554  \\
				DDMEX1~&0.1506 ~&-16.0366~&291.8938~& 938.0250 ~&31.8400 ~&53.4410 ~& -66.8187 ~& 707.2811 ~&0.5709 ~&0.5709   \\
				DDMEX2~&0.1521 ~&-16.0376~&255.7117~& 631.8805 ~&35.3064 ~&86.8583 ~& -55.3598 ~&  99.1595 ~&0.5780 ~&0.5780   \\
				DDMEXY~&0.1536 ~&-16.0241~&268.6004~& 783.4160 ~&32.0443 ~&53.2243 ~& -74.2635 ~& 736.0733 ~&0.5811 ~&0.5811   \\
				DDME2 ~&0.1521 ~&-16.1417~&251.8909~& 480.5739 ~&32.3178 ~&51.2763 ~& -87.1828 ~& 776.7706 ~&0.5718 ~&0.5718  \\
				DDME1 ~&0.1521 ~&-16.2327~&244.4246~& 317.6665 ~&33.0687 ~&55.4385 ~&-101.0120 ~& 706.4165 ~&0.5776 ~&0.5776  \\
				DD2   ~&0.1491 ~&-16.6679~&241.4682~& 169.3215 ~&31.6642 ~&55.0275 ~& -93.0319 ~& 599.0228 ~&0.5627 ~&0.5614  \\
				PKDD   ~&0.1496 ~&-16.9145~&261.3327~&-118.5887 ~&36.7950 ~&90.2315 ~& -80.4556 ~&  24.7448 ~&0.5713 ~&0.5699  \\
				TW99  ~&0.1531 ~&-16.2471~&240.5921~&-539.9163 ~&32.7742 ~&55.3160 ~&-124.7304 ~& 538.8057 ~&0.5549 ~&0.5549  \\
				DDV   ~&0.1516 ~&-16.9278~&241.1875~&-612.3358 ~&33.6850 ~&69.8717 ~& -97.7363 ~& 259.8839 ~&0.5869 ~&0.5852  \\
				DDVT  ~&0.1536 ~&-16.9155~&239.1848~&-743.0882 ~&31.5692 ~&42.3958 ~&-118.8231 ~& 896.2710 ~&0.6670 ~&0.6657  \\
				DDVTD  ~&0.1535 ~&-16.9165~&239.3400~&-762.4664 ~&31.8052 ~&42.5800 ~&-117.3942 ~& 873.0518 ~&0.6673 ~&0.6660  \\ 
				DDT   ~&0.1551 ~&-16.8727~&229.2688~&  90.7116 ~&31.2197 ~&34.2543 ~& -69.2275 ~&1208.8295 ~&0.6574 ~&0.6560  \\
				\hline\hline    
		\end{tabular}
\end{table}

\section{Results and discussions}
\begin{figure}[htbp]
		\includegraphics[scale=0.7]{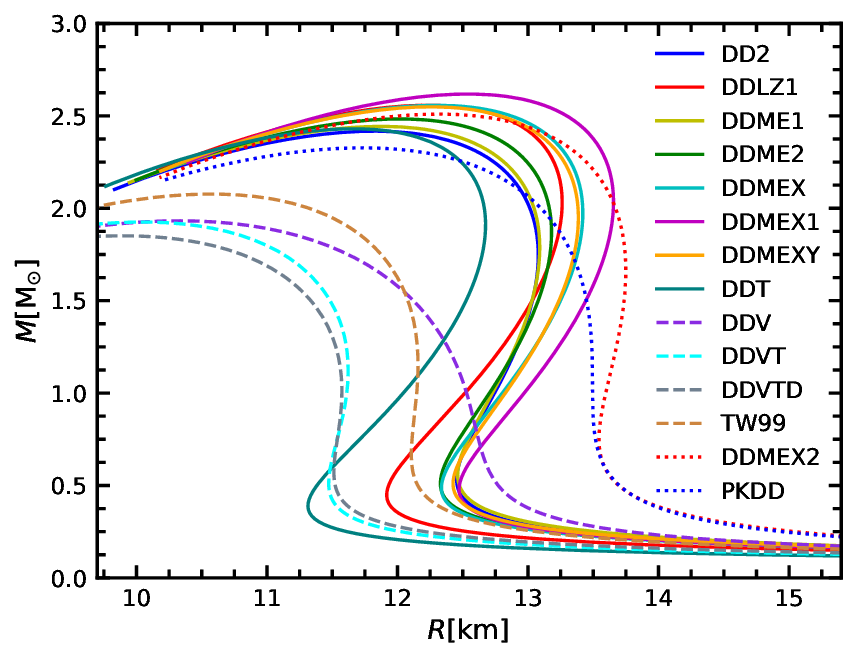}
		\caption{\label{fig1}The mass-radius relations of NS from 14 DDRMF parameter sets.}
	\end{figure}

	Initially, the mass-radius ($M$-$R$) relations of NSs based on EOSs with 14 DDRMF parameterizations are illustrated in Fig.~\ref{fig1}, where the crust EOS is derived from the original TM1 parameter set. The maximum masses of NSs predicted by these EOSs range from approximately $1.9M_\odot$ to $2.6M_\odot$, consistent with recent astronomical observations \citep{Huang2022}. These EOSs can be classified into three types:
	
	1. Low-Mass Group: Represented by dashed lines, this group includes EOSs with maximum masses below $2.0M_\odot$ and smaller radii at $1.4M_\odot$. The skewness coefficient of the binding energy per nucleon, $Q$, is notably low in this group.
	
	2. Intermediate-Mass Group: EOSs in this category, shown with solid lines, feature maximum masses exceeding $2.0M_\odot$ and radii at $1.4M_\odot$ within an intermediate range.
	
\begin{figure}[htbp]
	 	\centering
		\includegraphics[scale=0.7]{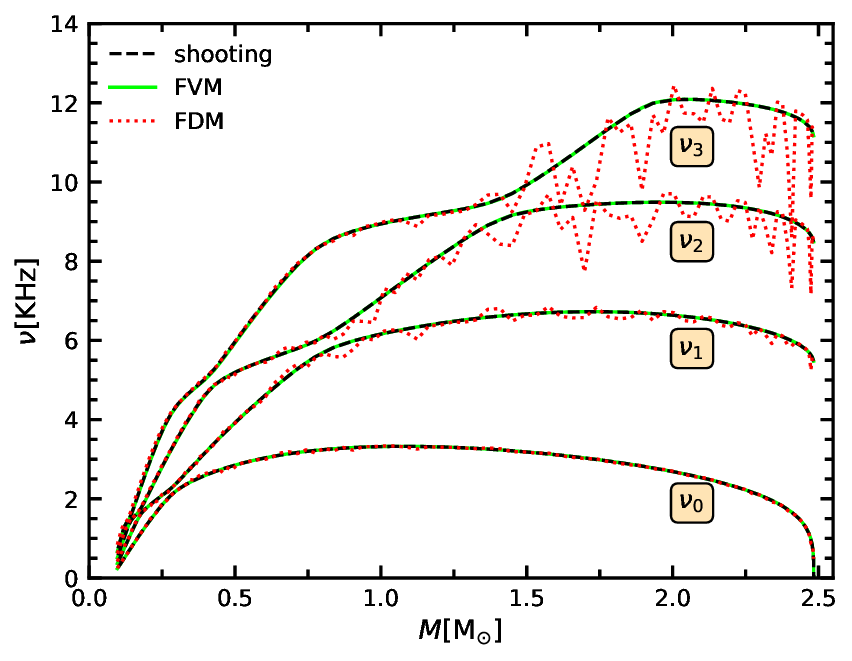}
		\caption{\label{fig2}The first four eigenfrequencies of the radial oscillation as functions of NS masses with the EOS obtained by DDME2 parameter set from shooting method, FDM, and FVM.}
	\end{figure}

	3. High-Radius Group: Indicated by dotted lines, this group includes EOSs with the largest radii at $1.4M_\odot$. These EOSs exhibit significantly higher values of symmetry energy and its slope compared to the other parameter sets, where the $Q_{\rm sym}$ are also smaller.
	In Fig. \ref{fig2}, the first four eigenfrequencies of NS radial oscillations are presented for the EOS derived from the DDME2 parameterization, solved by the shooting method, FDM, and FVM. The results from the shooting method and FDM are in complete agreement; however, notable discrepancies arise between these methods and FVM, particularly for higher-order modes at larger NS masses. These differences stem from the presence of the crust, where the nodes of higher-order oscillations are primarily concentrated.  {The discontinuities of the speed of sound within the crust region significantly affect the accuracy of the FDM, which leads to the observed fluctuations in oscillation frequencies,  since the speed of sound, $c^2_s$ in $H$ function in Eq. (\ref{ode}) must be differentiated. Additionally, as the frequency mode increases, the influence of the crust becomes more prominent, and the fluctuations in the results of the FDM become greater.}
	
	For the first four eigenfrequencies, the error between the FVM and the shooting method is less than $0.1\%$ when using $N=3000$ grid points, and the differences for the first 15 eigenfrequencies remain below $1\%$. Even with a reduced grid number, $N=100$, the fundamental mode ($\nu_0$) from FVM remains consistent with the shooting method. The frequency behavior across these modes as a function of NS mass aligns well with previously reported results, such as those in Refs. \citep{Apurba2023,Sen2023}.
	
\begin{figure}[htbp]
		\centering
		\includegraphics[scale=0.7]{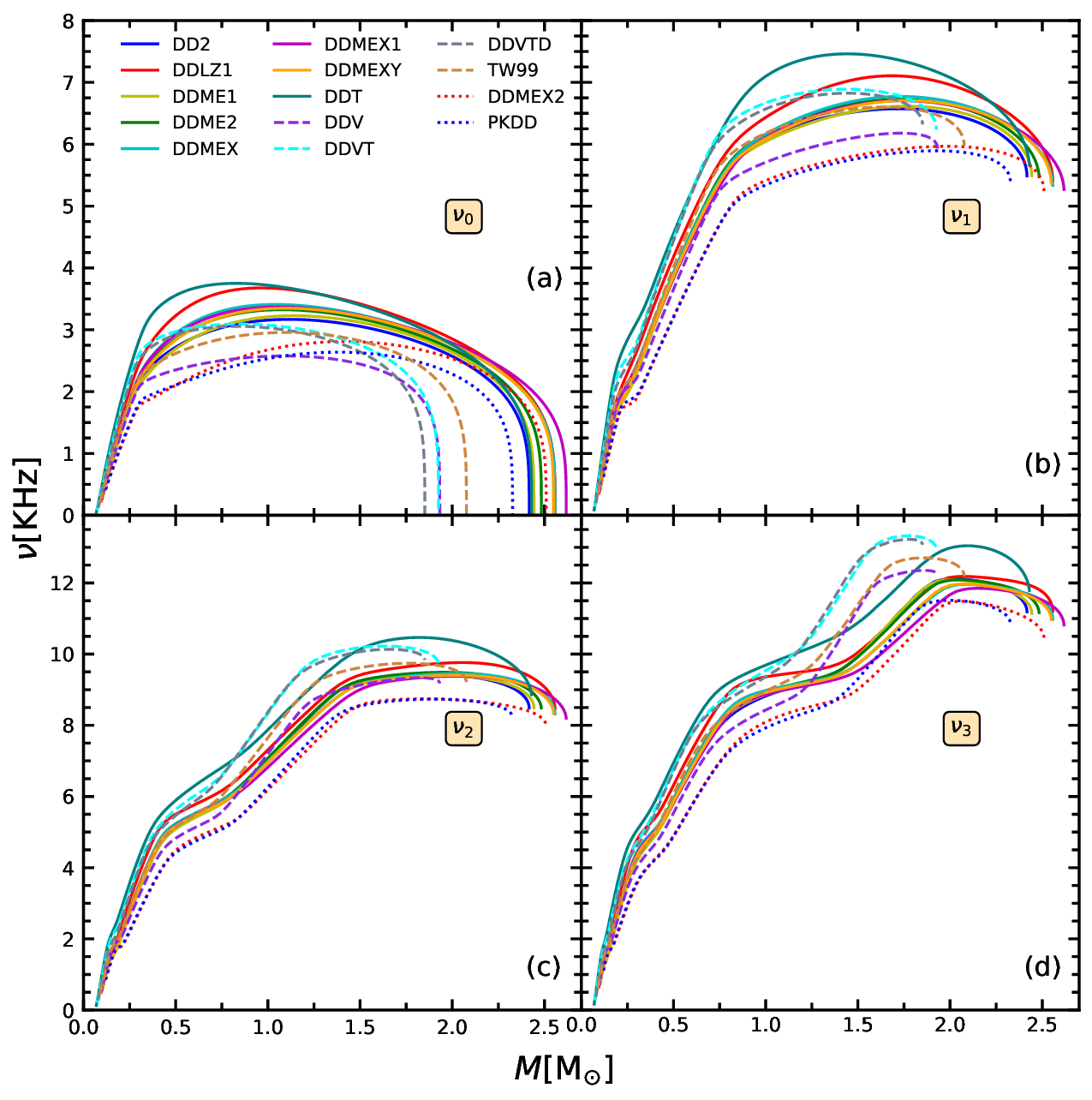}
		\caption{The first four frequencies of the radial oscillation as functions of NS masses obtained with the EOSs from different DDRMF parameterizations.}\label{fig3} 
	\end{figure}

	The frequencies of the first four modes as functions of NS mass, based on the EOSs from 14 DDRMF parameterizations, are shown in Fig. \ref{fig3}. These frequencies generally increase with NS masses in the lower mass region. For the fundamental mode ($\nu_0$) at panel (a), the frequency reaches a maximum value below $1.0M_\odot$ and then decreases in the higher mass region for EOSs with smaller symmetry energy. In contrast, for EOSs with larger symmetry energy ($E_{\rm sym}$) and slope of symmetry energy ($L$), such as DDMEX2 and PKDD, the maximum value of $\nu_0$ occurs around $1.5M_\odot$. {EOSs with relatively large masses exhibit different variations under different frequency modes.} At the maximum NS mass, $\nu_0$ becomes zero, satisfying the stability condition of the NS. In contrast, the frequencies of higher modes ($\nu_1$, $\nu_2$, and $\nu_3$) continue to increase with NS mass. These frequencies exhibit a strong correlation with the density-dependent behavior of the symmetry energy in the EOSs, a topic that will be discussed in more detail later.
	
\begin{figure}[htbp]
	\centering
		\includegraphics[scale=0.7]{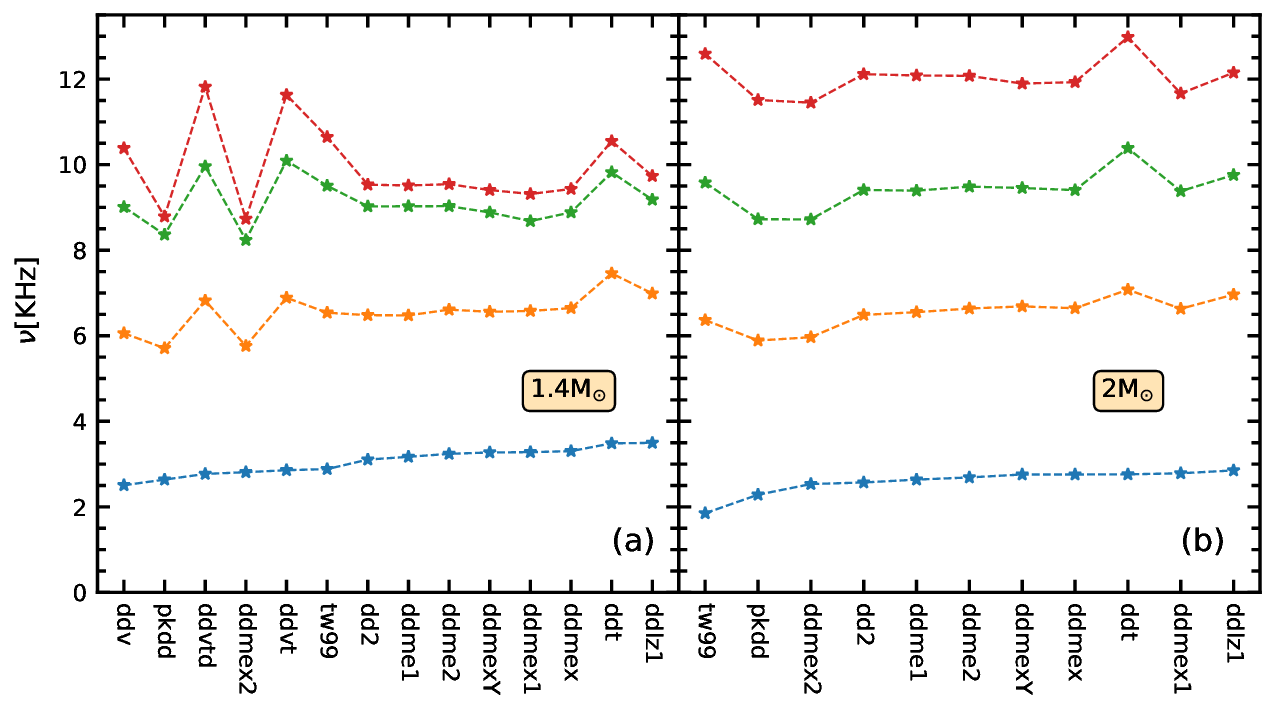}
		\caption{ The first four frequencies from the DDRMF EOSs at $1.4M_{\odot }$ ((a) panel) and  $2M_{\odot }$ are shown ((b) panel).}\label{fig4}
	\end{figure}

	The first four lowest frequencies at $1.4M_{\odot}$ and $2M_{\odot}$ for the current DDRMF EOSs are shown at panel (a) and panel (b) in Fig. \ref{fig4}, respectively. We begin by listing the fundamental frequencies from smallest to largest across the different DDRMF parameterizations. However, the higher-order oscillation modes do not display the consistent growth trend as the fundamental mode. Notably, for the higher frequencies at $1.4M_\odot$, the softer EOSs tend to have larger frequency differences. Meanwhile DDMEX2, PKDD and DDT, which have a very special skewness in the symmetry energy, also exhibit a staggering behavior. Furthermore, the frequency evolutions for $\nu_1$, $\nu_2$, and $\nu_3$ at $2M_{\odot}$ are consistent with each other, and these frequencies do not need to approach zero at the maximum NS mass.
	
\begin{figure}[htbp]
		\centering
		\includegraphics[scale=0.7]{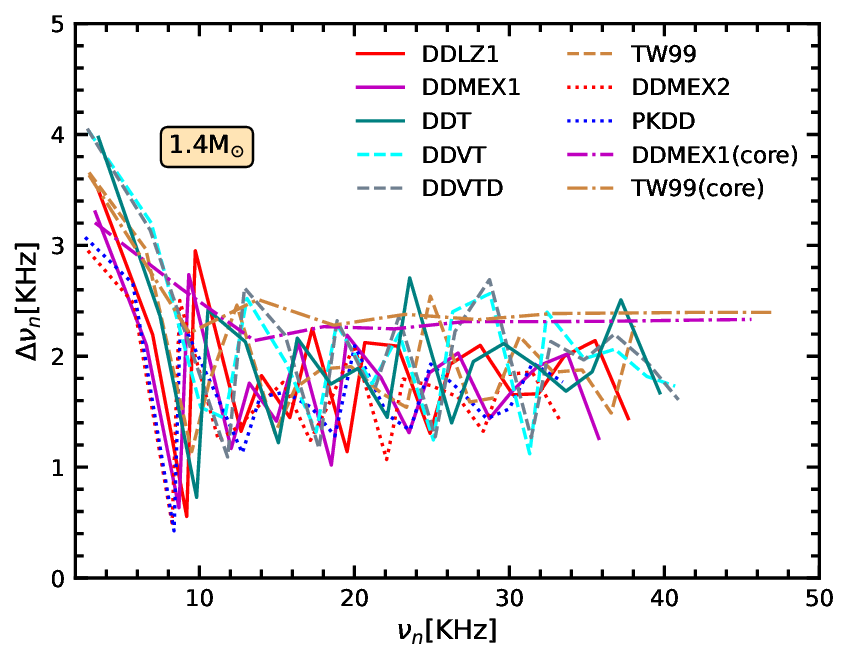}
		\caption{The relationships between oscillation frequencies $\nu_n$ and their differences $\Delta \nu_n=\nu_{n+1}-\nu_n$ at $1.4M_{\odot }$.}\label{fig5}
	\end{figure}

	The frequency difference, $\Delta \nu_n = \nu_{n+1} - \nu_n$, at $1.4M_{\odot}$ as a function of frequency $\nu_n$ for partial DDRMF parameterizations is shown in Fig. \ref{fig5}. {Compared with the EOSs without crust, the EOSs with crust show that $\Delta\nu_n$ varies unevenly with $\nu_n$.} This discrepancy arises because some oscillation nodes are located in the NS crust at higher-order modes. The non-uniform pasta structure in the crust affects the oscillation frequency due to the non-monotonic behaviour of the adiabatic index.  {The distribution of the first large separation $\Delta\nu_0$ is roughly opposite to the corresponding radius distribution, and the third large separation $\Delta\nu_2$ is generally the smallest among them.} The formation of a new node in the crust region typically corresponds to a peak in $\Delta \nu_n$. In general, the frequency difference decreases for higher-order oscillation modes, as there are many nodes in the crust region that have a weaker influence on the frequencies.
	
\begin{figure}[htbp]
	\centering
		\includegraphics[scale=0.7]{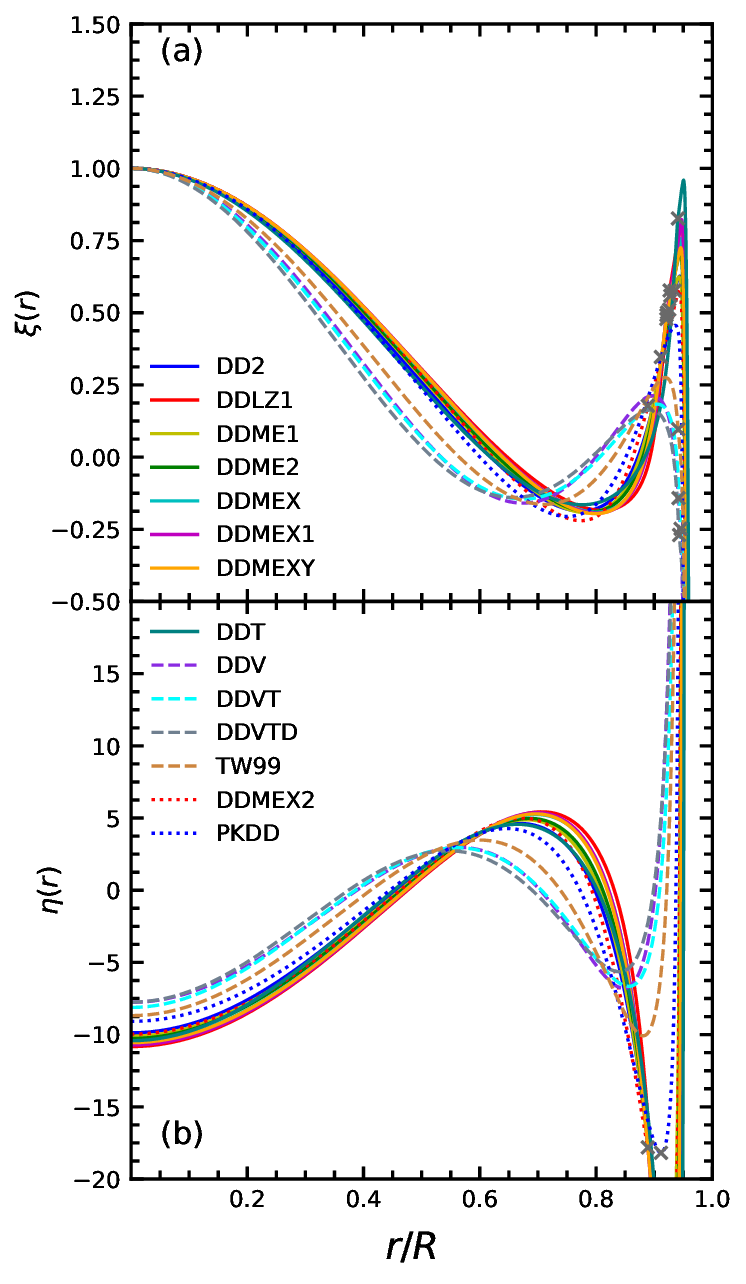}
		\caption{The radial displacement ((a) panel) and pressure perturbation ((b) panel) as functions of NS radius for NS with $1.4M_{\odot }$.}	\label{fig6} 
	\end{figure}

	The radial displacement and pressure perturbation, $\xi(r)$ and $\eta(r)$ for the $\nu_3$ mode as functions of radius at $1.4M_{\odot}$, are plotted at panel (a) and panel (b) in Fig. \ref{fig6}, respectively. The cross symbols represent the NS radii at the crust-core transition. They are the eigenfunctions of Eq.~\eqref{roeq}, with the nodes corresponding to the mode number. Both $\xi(r)$ and $\eta(r)$ exhibit more pronounced oscillations for higher modes. Their magnitudes undergo rapid changes between $0.9R$ and $R$, particularly after the crust-core phase transition, with this effect being more prominent for harder EOSs. The last nodes are located in the NS crust region, as discussed in previous studies \citep{Sun2021, Apurba2023}. Additionally, at $r \sim 0.7R$, $\xi$ and at $r \sim 0.55R$, $\eta$ from different EOSs have similar values.
	
\begin{figure}[htbp]
		\includegraphics[scale=0.7]{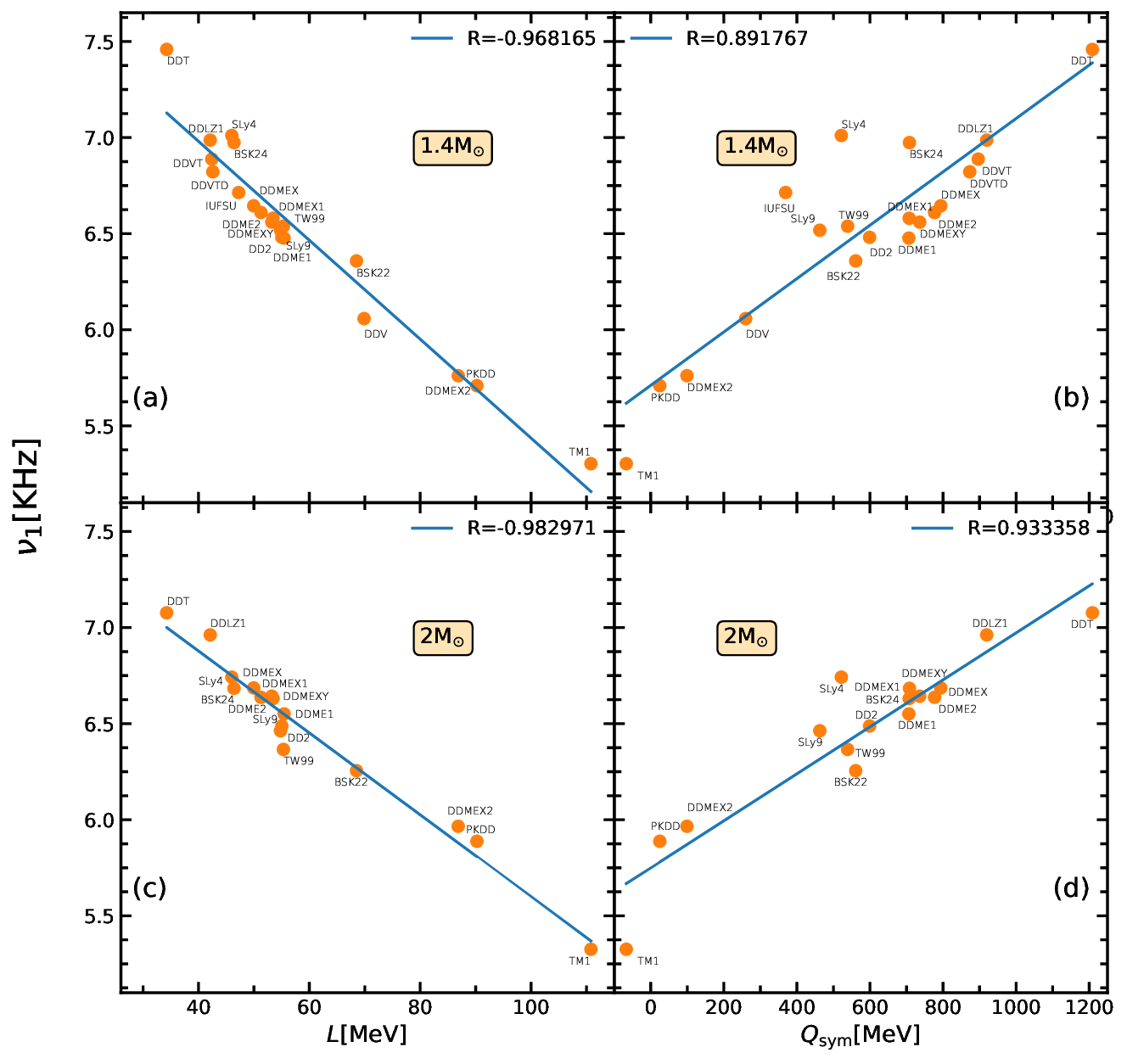}
		\caption{\label{fig7}The correlations between the frequencies $\nu_{1}$ and the slope $L$ and skewness coefficient $Q_{\rm{sym}}$ of symmetric energy at $1.4M_{\odot }$ ((a, b) panels) and $2M_{\odot }$ ((c, d) panels).}  
	\end{figure}

\begin{figure}[htbp]
		\centering
		\includegraphics[scale=0.7]{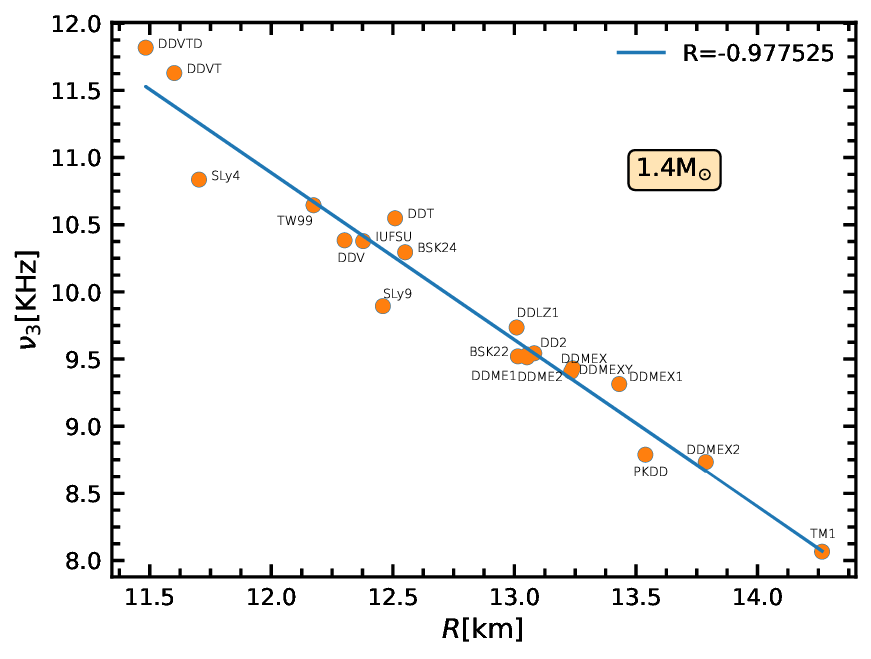}
		\caption{The relationship between NS radius and oscillation frequency $\nu_{3}$ at $1.4M_\odot$.}\label{fig8}
	\end{figure}

	Finally, the correlations between the nuclear saturation properties and oscillation frequencies are investigated using the EOSs from the DDRMF parameter sets. {Meanwhile, Skyrme-like models (SLy4, SLy9\citep{Paweł2009,Gulminelli2015,CHABANAT1998}, BSK22, BSK24\citep{Welker2017,Perot2019,Goriely2013,Pearson2019,Pearson2020,Pearson2022,Allard2021,Xu2013,Audi_2017}) and non-linear relativistic mean-field (NL-RMF) models (TM1\citep{Sugahara1994,Shen1998,Xia2022} and IUFSU\citep{Fattoyev2010}) were added for comparison.} It is found that the slope of symmetry energy, $L$, and the skewness coefficient of symmetry energy, $Q_{\rm sym}$, exhibit a strong linear correlation with the oscillation frequencies of the first excitation mode, $\nu_1$, at both $1.4 M_\odot$ and $2.0 M_\odot$, as shown in Fig.   \ref{fig7}. For the slope of symmetry energy, the correlation coefficients with the frequencies are negative, with {$R_{1.4} = -0.9681$ and $R_{2.0} = -0.9829$}, respectively. The EOSs with larger values of $L$ from the DDMEX2 and PKDD sets yield lower frequencies around $5.8$ kHz, while for the smallest $L$ from the DDT set, the frequency is around $7.0$ kHz. This suggests that oscillation frequencies could serve as a new probe to constrain the slope of symmetry energy through NS observations. On the other hand, $Q_{\rm sym}$ has a positive correlation with $\nu_1$, with correlation coefficients of {$R_{1.4} = 0.8917$ and $R_{2.0} = 0.9335$.} {We also  notice that the results of SLy4 and IUFSU models are a little deviation from the inverse linear relation.} $Q_{\rm sym}$ corresponds to the third term of the density dependence in symmetry energy, which becomes dominant in the high-density region, but cannot be efficiently measured with current nuclear experiments.
	Furthermore, the NS radius exhibits a strong negative linear dependence on the oscillation frequency of the third excitation mode, $\nu_3$, at $1.4M_\odot$ for the DDRMF models, with a correlation coefficient of {$R = -0.9775$} as given in Fig. \ref{fig8}. {Among all the DDRMF parameters, DDMEX2 set generates the largest NS radius at $1.4M_\odot$,} $13.8$ km, and the lowest oscillation frequency, $\nu_3 = 8.7$ kHz, while the EOS from the DDVTD set produces the smallest NS radius, $11.5$ km, and the highest oscillation frequency, $\nu_3 = 11.8$ kHz. This strong correlation suggests that it is possible to determine the radius of a NS by detecting its radial oscillation frequency at higher modes.

\section{Summary and perspectives}\label{IV}
The radial oscillation of NSs was systematically investigated using EOSs from DDRMF parameter sets, which accurately describe the ground-state properties of finite nuclei. We examined the impact of the conservative form on the radial oscillation problem of NS with discontinuous coefficients and introduced the FVM. {This approach preserves the conservation properties of the difference equations and yields accurate frequency calculations.}
	
	The EOSs from the 14 DDRMF parameterizations were used to calculate NS oscillation frequencies and the radial displacement and pressure perturbation, $\xi$ and $\eta$, for various modes, from the fundamental mode to high-order modes. The variation tendencies for the high-order modes were consistent across the EOSs. Frequency differences varied unevenly with oscillation frequencies, and rapid changes in $\xi$ and $\eta$ were observed in the crust region for higher modes, where nodes of the eigenfunctions appear in the crust.
	
	Correlations between the saturation properties of symmetric nuclear matter and oscillation frequencies were examined for the DDRMF EOSs, { as well as six EOSs including Skyrme-like and NL-RMF models.} {Strong linear correlations were found between the slope of symmetry energy, $L$ and the oscillation frequency of the first excitation mode, $\nu_1$, at $1.4M_\odot$ and $2.0M_\odot$, respectively. Larger values of $L$ correspond to smaller values of $\nu_1$. Similarly, there is an inverse linear relationship between the skewness coefficient of symmetry energy, $Q_{sym}$, and $\nu_1$.} Additionally, the NS radius at $1.4M_\odot$ showed a strong negative correlation with the oscillation frequency of the third excitation mode, $\nu_3$. {The EOSs of Skyrme-like and NL-RMF models exhibit the same properties as those of the DDRMF model, which indicates the universality of these relationships.} These correlations provide new probes for constraining the density-dependent behavior of symmetry energy and suggest a method for measuring the NS radius by detecting its high-order radial oscillation frequencies. Radial oscillations of NSs alone cannot generate GWs. However, they can couple with non-radial oscillations of NSs, amplifying GWs and increasing their detectability. Although current GW detectors are unable to observe GWs in this frequency range, third-generation detectors, such as the Einstein Telescope, are designed to achieve a detection bandwidth spanning from 1 Hz to 10 kHz \citep{Punturo2010,Hild2010}. Additionally, the Cosmic Explorer is expected to bring significant advancements in GW detection. Radial oscillations of NSs can also modulate short gamma-ray bursts (SGRBs), which may originate from the merger of two NSs. Therefore, detecting SGRBs could provide a means of confirming the oscillation frequencies. A promising detection approach, which utilizes the modulation of SGRBs and requires lower detector sensitivity, has been proposed \cite{Chirenti2019}. However, it is still likely that third-generation GW detectors will be necessary to observe the target frequencies. Thus, it is reasonable to expect that third-generation GW detectors will validate these conclusions.
    
    Future work will focus on non-radial oscillations. However, when addressing non-radial oscillation problems, the unique structure of the equation prevents isolating the frequency on one side and reformulating it as an eigenvalue problem in matrix form. Therefore, we are actively exploring potential methods to fully leverage the capabilities of FVM.

\vspace{-1mm}
\centerline{\rule{80mm}{0.1pt}}
\vspace{2mm}

\bibliographystyle{apsrev4-1}
\bibliography{ref}

\end{document}